\documentclass[12pt]{iopart}
\input{amssym}
\usepackage{epsfig}
\global\arraycolsep=2pt

\newcommand{\be}{\begin{equation}}

\newcommand{\ee}{\end{equation}}
\newcommand{\bea}{\begin{eqnarray}}
\newcommand{\eea}{\end{eqnarray}}
\newcommand{\ben}{\begin{equation*}}
\newcommand{\een}{\end{equation*}}
\newcommand{\bean}{\begin{eqnarray*}}
\newcommand{\eean}{\end{eqnarray*}}

\begin{document}

\title{Exact zero-point interaction energy between cylinders}

\author{F.D. Mazzitelli$^{(a)}$, D.A.R. Dalvit$^{(b)}$, and F.C. Lombardo$^{(a)}$ }

\address{$^{(a)}$Departamento de F\'{\i}sica J.J. Giambiagi, Facultad de Ciencias Exactas y
Naturales, Universidad de Buenos Aires - Ciudad Universitaria, Pabell\'on I, 1428 Buenos Aires, Argentina}

\address{$^{(b)}$Theoretical Division, MS B213, Los Alamos National Laboratory, Los Alamos, NM 87545, USA}

\date{\today}

\begin{abstract}
We calculate the exact Casimir interaction energy between two
perfectly conducting, very long, eccentric cylindrical shells
using a mode summation technique. Several limiting cases of the
exact formula for the Casimir energy corresponding to this
configuration are studied both analytically and numerically. These
include concentric cylinders, cylinder-plane, and eccentric
cylinders, for small and large separations between the surfaces.
For small separations we recover the proximity approximation,
while for large separations we find a weak logarithmic decay of the Casimir
interaction energy, typical of cylindrical geometries.
\end{abstract}


\maketitle

\section{Introduction}

Almost 60 years ago \cite{Casimir1948}, Casimir discovered an
interesting macroscopic consequence of the zero point fluctuations
of the electromagnetic field: an attractive force between
uncharged parallel conducting plates. Since then, the dependence
of the Casimir force with the geometry of the conducting surfaces
has been the subject of several works \cite{reviews}. For many
years, the only practical way to compute the Casimir energy for
non planar configurations was the so called proximity force
approximation (PFA) \cite{Derjaguin1957}. This approximation is
valid for surfaces whose separation is much smaller than typical
local curvatures. Due to the high precision experiments performed
since 1997 \cite{exp}, there has been a renewed interest in the
geometry dependence of the Casimir force, and in particular in the
calculations of the corrections to the proximity approximation.

In the last years there have been a number of attempts to compute
the Casimir forces beyond the PFA, using for example semiclassical
\cite{sem, Mazzitelli2003} and optical \cite{opt} approximations, and
numerical path-integral methods  \cite{num}.
Large deviations from PFA for corrugated plates have been reported \cite{Genet2003},
and in recent months the Casimir
energy has been computed exactly for several configurations of
experimental interest, as the case of a sphere in front of a
plane, and a cylinder in front of a plane
\cite{Bulgac2006,Emig2006,Bordag2006,Gies2006}. As first suggested
in \cite{Dalvit2004}, the latter configuration is intermediate
between the sphere-plane and the plane-plane geometries, and may
shed light on the longstanding controversy about thermal
corrections to the Casimir force. There is an ongoing experiment
to measure precisely the Casimir force for this geometry
\cite{Brown-Hayes2005}.

The configuration of two eccentric cylinders is of experimental relevance too
\cite{Dalvit2004,Mazzitelliqftext,Dalvit2006}. Although parallelism
is as difficult as for the plane-plane configuration, the fact
that the concentric configuration is an unstable equilibrium
position opens the possibility of measuring the derivative of the
force using null experiments. Up to now, the Casimir interaction
energy between two cylindrical shells has been computed
semiclassically and exactly in the concentric case
\cite{Mazzitelli2003,Saharian2006}, and using the proximity
approximation in the eccentric situation
\cite{Dalvit2004,Mazzitelliqftext}. In principle, one could
consider experimental configurations in which a very thin metallic
wire is placed inside a larger hollow cylinder. In this case, a
more accurate determination of the Casimir force is needed. The
aim of this paper is to describe in detail the derivation of the exact Casimir
interaction energy for eccentric cylinders, initially reported by us in \cite{Dalvit2006},
and to compute analytically different limiting cases of relevance for Casimir force
measurements in this configuration. To this end we will
use the mode summation technique combined with the argument
theorem in order to write the Casimir energy as a contour integral
in the complex plane \cite{Nesterenko1998}.

The paper is organized as follows. In Section 2 we derive an
expression for the Casimir interaction energy for any
configuration invariant under translations in one of the spatial
dimensions. When properly subtracted, this expression reduces to an
integral over the imaginary axis, and is similar to expressions
for the Casimir energy derived using path integrals or scattering
methods. In Section 3 we derive the exact formula for the
interaction energy between eccentric cylinders and we
analyze some particular cases of the exact formula. We first show
the known results for the concentric case obtained from
the exact formulation, and that it is possible to derive the interaction energy for the
cylinder-plane configuration in the appropriate limit. In Section 4 we
consider the exact formula in the limit of quasi concentric cylinders
of arbitrary radii. We discuss two opposite limits of
this exact formula: large and small separations between the
metallic surfaces. In the first limit, we find that the Casimir
energy between a thin wire contained in a hollow cylinder has a
weak logarithmic decay as the ratio between the outer and inner radii becomes
very large. In the second limit, we recover previous results
obtained using PFA for quasi concentric cylinders.
Finally, Section 5 contains the conclusions of our work.


\section{Casimir energy as a contour integral}

The  Casimir energy for a system of conducting shells can be
written as
\begin{equation}
E_c= \frac{1}{2} \sum_p(w_p-\tilde w_p) ,
\label{ecasmodes}
\end{equation}
where $w_p$ are the eigenfrequencies of the electromagnetic field
satisfying perfect conductor boundary conditions on the surfaces
of the conductors, and $\tilde w_p$ are those corresponding to the
reference vacuum (conductors at infinite separation). Throughout this
paper we use units $\hbar=c=1$. The subindex
$p$ denotes the set of quantum numbers associated to each
eigenfrequency. Introducing a cutoff for high frequency modes
$E_{c}(\sigma)={1\over 2}\sum_p(e^{-\sigma w_p} w_p-e^{-\sigma
\tilde w_p} \tilde w_p)$,
the Casimir energy $E_{c}$ is the limit of $E_{c}(\sigma)$ as
$\sigma\rightarrow 0$. For simplicity we choose an exponential
cutoff, although the explicit form is not relevant.

Let us consider a general geometry with translational invariance
along the $z-$axis (as for example very long and parallel waveguides
of arbitrary sections). The transverse electric (TE) and
transverse magnetic (TM) modes can be described in terms of two
scalar fields with adequate boundary conditions. In cylindrical
coordinates, the modes of each scalar field will be of the form
$h_{n, k_z}(t,r,\theta,z)=e^{(-iw_{n, k_z}t+ik_z
z)}R_n(r,\theta)$, where the eigenfrequencies are $w_{n,
k_z}=\sqrt{k_z^2+\lambda^2_n}$, and $\lambda_n$ are the
eigenvalues of the two dimensional Laplacian
\begin{equation}
\left(\frac{\partial^2}{\partial
r^2}+\frac{1}{r}\frac{\partial}{\partial r}+
\frac{1}{r^2}\frac{\partial^2}{\partial\theta^2}+\lambda_n^2\right
) R_n(r,\theta)=0.
\end{equation}
The set of quantum numbers $p$ is given by
$(n, k_z)$. For very long cylinders of length $L$ we can replace
the sum over $k_z$ by an integral. The result is
\begin{equation}
E_{c} (\sigma) = {L \over 2}\int_{-\infty}^{\infty}{dk_z\over
2\pi} \sum_{n}\left (\sqrt{k_z^2+\lambda_{n}^2}e^{-\sigma
\sqrt{k_z^2+\lambda_{n}^2}} - \sqrt{k_z^2+\tilde \lambda_{n}^2}
e^{-\sigma \sqrt{k_z^2+\tilde\lambda_{n}^2}} \right) \; .
\label{exs}
\end{equation}
From the argument theorem it follows that
\begin{equation}
{1\over 2\pi i} \int_{C} \,d\lambda \;  \lambda \; e^{-\sigma \lambda} {d\over d\lambda}
\ln f(\lambda)=\sum_i \lambda_i \; e^{-\sigma \lambda_i} \; ,
\end{equation}
where $f(\lambda)$ is an analytic function in the complex $\lambda$ plane within the closed contour
${C}$, with simple zeros at $\lambda_1, \lambda_2, \dots$ within ${C}$.
We use this result to replace the sum over $n$ in Eq.(\ref{exs}) by a contour integral
\begin{equation}
E_c (\sigma)={L\over 4\pi i}\int_{-\infty}^{\infty} {dk_z\over 2\pi}
\int_{C} d\lambda \sqrt{k_z^2+\lambda^2}
e^{-\sigma \sqrt{k_z^2+ \lambda^2}}{d\over d\lambda}
\ln  \left( \frac{F}{F_{\infty}} \right)  ,
\end{equation}
where $F$ is a function that vanishes at $\lambda_n$ for all $n$
(and $F_{\infty}$ vanishes at $\tilde\lambda_n$).

\begin{figure}
\centerline{\psfig{figure=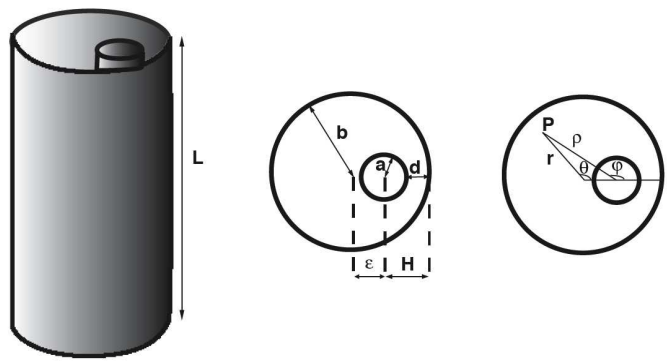,height=5cm,width=9cm,angle=0}}
\caption{Geometrical configuration studied in this paper. Two perfectly conducting eccentric
cylinders of radii $a<b$, length $L$, and eccentricity $\epsilon$ interact via the Casimir force.
The figure on the right shows the polar coordinates $(r,\theta)$ and $(\rho,\varphi)$ of any point $P$
between the eccentric cylinders used for the determination of the classical eigenvalues for this
configuration.}
\label{fig1}
\end{figure}

In the rest of this section we will consider the particular
configuration of two eccentric cylinders with circular sections of
radii $a$ and $b$, respectively. We will denote the eccentricity
of the configuration by $\epsilon$ (see Fig. 1).
The geometrical dimensionless parameters $\alpha \equiv  b/a$
and $\delta \equiv  \epsilon/a$ fully characterize the eccentric cylinder configuration.
It is worth emphasizing that the results of this section can be
trivially extended to more general configurations, as long as they
are translationally  invariant along one spatial dimension. It is
convenient to compute the difference between the energy of the
system of two eccentric cylinders and the energy of two isolated
cylindrical shells of radii $a$ and $b$,
\begin{equation}
E_{12}(\sigma)=E_c(\sigma) - E_1(\sigma,a)-E_1( \sigma,b) \; ,
\end{equation}
where
\begin{equation}
E_1(\sigma,a) ={L\over 4\pi i} \int_{-\infty}^{\infty} {dk_z\over 2\pi}
\int_{C} d\lambda \sqrt{k_z^2+ \lambda^2}
e^{-\sigma \sqrt{k_z^2+\lambda^2}}{d\over d\lambda}
\ln \left( \frac {F_{1{\rm cyl}}(a)}{F_{1{\rm cyl}}(\infty)} \right) \; .
\end{equation}
Here $F_{1{\rm cyl}}(a)$ is a function that vanishes at the
eigenfrequencies of an isolated cylindrical shell of radius $a$.
Therefore
\begin{equation}
E_{12}(\sigma) = {L\over 4\pi i}\int_{-\infty}^{\infty} {dk_z\over 2\pi}
\int_{C} d\lambda \sqrt{k_z^2+\lambda^2}
e^{-\sigma \sqrt{k_z^2+\lambda^2}}{d\over d\lambda} \ln M(\lambda) \; ,
\end{equation}
where
\begin{equation}
M= \frac{F}{F_{\infty}} \frac{F_{1{\rm cyl}}(\infty)^2} {F_{1{\rm cyl}}(a)F_{1{\rm cyl}}(b)} \; .
\label{Mformal}
\end{equation}

\begin{figure}
\centerline{\psfig{figure=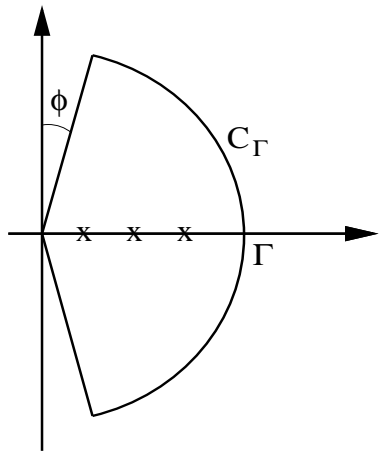,height=6cm,width=5cm,angle=0}}
\caption{Contour for the integration in the complex plane.}
\label{fig2}
\end{figure}

To proceed we must choose a contour for the integration in the
complex plane. In order to compute $E_{c} (\sigma),
E_1(\sigma,a)$, and $E_1(\sigma,b)$ separately, an adequate
contour is a circular segment ${C}_{\Gamma}$ and two straight line
segments forming an angle $\phi$ and $\pi -\phi$ with respect to
the imaginary axis (see Fig. 2). The nonzero angle $\phi$ is
needed to show that the contribution of ${C}_{\Gamma}$ vanishes in
the limit $\Gamma\rightarrow \infty$ when $\sigma > 0$. For the
rest of the contour, the divergences in $E_{c}(\sigma)$ are
cancelled out by those of $E_1(\sigma,a)$ and $E_1(\sigma,b)$, as
in the case of concentric cylinders  \cite{Mazzitelli2003}.
Therefore, in order to compute $E_{12}(\sigma)$ we can set
$\phi=0$ and $\sigma =0$, and the contour integral reduces to an
integral on the imaginary axis. We find
\begin{equation}
E_{12}= -{L \over 2\pi}\int_{-\infty}^{\infty} {dk_z\over 2\pi} ~{\rm Im}
\left\{\int_0^{\infty} dy \sqrt{k_z^2-y^2} {d\over dy}\ln M(iy)
\right\} \; . \label{xx}
\end{equation}
As we will see, $M(iy)$ is a real function - hence,  the integral
over $y$ in Eq. (\ref{xx}) is restricted to $y > k_z$. After some straightforward
steps one can re-write this equation as
\begin{equation}
E_{12}={L\over 4\pi} \int_{0}^{\infty} dy \ y\ln M(iy)  \; .
\label{xxx}
\end{equation}
As we have already mentioned, a similar expression can be derived
for conductors of arbitrary shape, as long as there is
translational invariance along the $z$-axis. It is worth noting
that the structure of this expression is similar to the ones
derived recently for the cylinder-plane geometry using path
integrals \cite{Emig2006,Bordag2006}, and for the sphere-plane
geometry using the Krein formula \cite{Bulgac2006}.


\section{The exact formula}

In this section we derive the exact formula for the Casimir energy
between eccentric cylinders. We proceed in two steps: we first
find the function $F$ with zeroes at the eigenfrequencies for the
geometric configuration. Then we obtain an explicit expression for
the function $M$, which involves a definition of the Casimir energy as a difference
between the energy of the actual configuration and a configuration
with very large and separated conductors.


\subsection{The classical eigenvalues}

The solution of the Helmholtz equation in the annular region
between eccentric cylinders has been considered in the framework
of classical electrodynamics, fluid dynamics, and reactor physics,
among others \cite{Singh1984,Balseiro1950}. The eigenvalues have
been computed using different approaches, as for instance
conformal transformations that map the eccentric annulus onto a
concentric one. As the two dimensional Helmholtz equation is not
conformally invariant, the transformed equation has coordinate
dependent coefficients and has to be solved numerically
\cite{Hine1971}. As is well known, it is very difficult to compute
the Casimir energy from the numerical eigenfrequencies. It is more
efficient to use the procedure outlined in Section 2, that only
needs a function $F$ with zeroes at the eigenvalues. Although for
the eccentric annulus this function  has been previously found in
the literature \cite{Singh1984,Balseiro1950}, for the benefit of
the reader we include here a derivation of this result.

The electromagnetic field inside an eccentric waveguide can be
described in terms of TM and TE modes. The TM modes are
characterized by a vanishing $z$ component of the magnetic field,
$B_z=0$. The other components of the electromagnetic field can be
derived from the $z$ component of the electric field,
$E_z(r,\theta,z,t)=E(r,\theta) e^{-i \omega t + i z k_z}$, with
\begin{equation}
E(r,\theta)=\sum_m\left[A_m J_m(\lambda r)+B_m N_m(\lambda
r)\right]e^{im\theta} ,
\label{E}
\end{equation}
where $w^2=k_z^2+\lambda^2$, and $(r,\theta)$ are polar
coordinates with origin at the center of the outer cylinder (see Fig. 1).
One can also describe the $z$ component of the electric field
using polar coordinates $(\rho , \varphi) $
with origin at the center of the inner cylinder (see Fig. 1),
\begin{equation}
\bar E(\rho,\varphi)=\sum_n\left[\bar A_n J_n(\lambda \rho)+\bar
B_n N_n(\lambda \rho)\right]e^{in\varphi}\, . \label{barE}
\end{equation}
The perfect conductor boundary conditions imply that the
$z$ component of the electric field must vanish on the cylindrical
shells
\begin{eqnarray}
A_m J_m(\lambda b)+B_m N_m(\lambda b)&=& 0 ,\nonumber\\
\bar A_n J_n(\lambda a)+\bar B_n N_n(\lambda a) &=& 0\,\, ,
\label{ABdir}
\end{eqnarray}
i.e., the functions $E$ and $\bar E$ satisfy Dirichlet boundary
conditions on the surfaces. The coefficients of the series in Eqs.(\ref{E}) and (\ref{barE})
can be related to one another using the addition theorem for Bessel functions
\begin{equation}
e^{im\varphi}{\cal C}_m(\lambda\rho)=\sum_p e^{ip\theta}{\cal
C}_p(\lambda r)J_{p-m}(\lambda\epsilon)\,\, , \label{addition1}
\end{equation}
where ${\cal C}_m$ denotes either $J_m$ or  $N_m$. Indeed, as at any
point $P$ in the annulus region one must have $E(P)=\bar E(P)$, it
is possible to show that
\begin{eqnarray}
\bar A_n &=& \sum_m A_m J_{n-m}(\lambda\epsilon) , \nonumber\\
\bar B_n &=& \sum_m B_m J_{n-m}(\lambda\epsilon) .
\label{relAB}
\end{eqnarray}
Combining Eqs. (\ref{ABdir}) and (\ref{relAB}) one obtains the linear, homogeneous
system of equations
\begin{equation}
\sum_m A_m \left[
\frac{J_n(\lambda a)}{N_n(\lambda a)} - \frac{J_m(\lambda b)}{N_m(\lambda b)}
\right] J_{n-m}(\lambda \epsilon) = 0 .
\end{equation}
The solution of this linear system of equations is non trivial only if
${\rm det} [ Q^{\rm TM}_{mn} ] =0$, where
\begin{equation}
Q^{\rm TM}_{mn}(a,b,\epsilon)= \left[J_n(\lambda a) N_m(\lambda b)
-J_m(\lambda b) N_n(\lambda a)\right] J_{n-m}(\lambda
\epsilon)\,\, .
\end{equation}
This equation defines the allowed values for $\lambda$, and
therefore defines the eigenfrequencies of the TM modes.

The TE modes can be treated in the same fashion. For these modes
the $z$ component of the electric field vanishes in the annulus
region, $E_z=0$. The perfect conductor boundary conditions imply
that the normal component of the magnetic field should vanish on
the conducting shells, so now we must impose Neumann boundary
conditions on the surfaces. The eigenvalues $\lambda$ for the TE
modes are the solutions of ${\rm det} [Q^{\rm TE}_{mn}]=0$, where
\begin{equation}
Q^{\rm TE}_{mn}(a,b,\epsilon)= \left[J'_n(\lambda a) N'_m(\lambda
b) -J'_m(\lambda b) N'_n(\lambda a)\right] J_{n-m}(\lambda
\epsilon).
\end{equation}

In the concentric limit $\epsilon =0$ we have
$J_{n-m}(0)=\delta_{nm}$, the two matrices  $Q^{\rm TE}_{mn}$ and
$Q^{\rm TM}_{mn}$ become diagonal, and the equations for the
eigenvalues are those of the concentric case
\cite{Mazzitelli2003}. In what follows we will use these matrices
to define the function $M$ that enters in Eq.(\ref{xxx}).


\subsection{The function $M$}

Roughly speaking, the function $M$ that determines the Casimir
energy through Eq.(\ref{xxx}) is the ratio of the function associated to
the actual geometric configuration and the one associated to a
configuration in which the conducting surfaces are very far away from each other.
As the last configuration is not univocally defined, we will use
this freedom to choose a particular one that simplifies the
calculation. It turns to be convenient to subtract a configuration
of two cylinders with very large (and very different) radii, but
with the same eccentricity as that of the configuration of interest.

We start considering the Dirichlet modes.
To compute $F_{1{\rm cyl}}(a)$ in Eq.(\ref{Mformal})  we note that the
eigenfrequencies  $\lambda$ for the geometry of a single cylinder of radius $a$
surrounded by a larger one of radius $R$ are defined by the equations
\begin{eqnarray}
&& J_n(\lambda a)=0\, ,\nonumber\\
&& J_n(\lambda a) N_n(\lambda R) - J_n(\lambda R) N_n(\lambda
a)=0\; . \label{condtm}
\end{eqnarray}
The first equation defines the eigenfrequencies in the region
$r<a$ and the second one gives the eigenfrequencies of the modes
in the region $a<r<R$. $F_{1{\rm cyl}}(a)$ is the product of these
two relations for all values of $n$,  evaluated on the imaginary
axis ($\lambda=i y\equiv i\beta/a$). Namely,
\begin{eqnarray}
F_{1{\rm cyl}}(a)&=&\prod_n J_n(\lambda a)[J_n(\lambda a)N_n(\lambda R)-J_n(\lambda R)
N_n(\lambda a)]\nonumber \\
& \equiv & J(a) {\rm det}[ Q^{\rm TM}(a,R,0)] \; ,
\end{eqnarray}
where we have introduced the notation  $J(a) \equiv \prod_n
J_n(\lambda a)$ to simplify the formulas below. The function
$F_{1{\rm cyl}}(\infty)$ has the same expression, but replacing
$a$ by $R_1$, with $R_1$ very large but smaller than $R$.  Using
the asymptotic expansion of the modified Bessel functions it is
easy to prove that $F_{1{\rm cyl}}(a) / F_{1{\rm cyl}}(\infty)
\simeq 2 \beta I_n(\beta) K_n(\beta) R_1/a$. The functions $F$ and
$F_{\infty}$ in Eq.(\ref{Mformal}) are given by
\begin{eqnarray}
F &=& J(a)  {\rm det}[Q^{\rm TM}(a,b,\epsilon)]   {\rm det}[ Q^{\rm TM}(b,R,0)]\nonumber \\
&=& \frac {J(a)}{J(b)}{\rm det}[Q^{\rm TM}(a,b,\epsilon)] F_{1{\rm cyl}}(b),
\label{efe} \\
F_{\infty} &=& \frac {J(R_1)}{J(R_2)}  {\rm det}[Q^{\rm
TM}(R_1,R_2,\epsilon)] F_{1{\rm cyl}}(\infty),
\label{efeinfinito}
\end{eqnarray}
where $R_1<R_2<R$. As we already stressed, in order to define $F_{\infty}$
we consider a configuration of two eccentric cylinders of large radii $R_1<R_2$ and with the same
eccentricity $\epsilon$ of the original configuration.
Evaluating the determinant in Eq.(\ref{efe})  on the imaginary axis one obtains
\begin{eqnarray}
{\rm det}[Q^{\rm TM}(a,b,\epsilon)] &=& {\rm det} \left[
\frac{2}{\pi} I_{n-m} \left( \beta\frac{\epsilon}{a} \right)
[K_n(\beta) I_m(\alpha\beta)\right. \nonumber \\
&-&\left. (-1)^{m+n}I_n(\beta)
K_m(\alpha\beta)] \right]. \label{detcomp}
\end{eqnarray}
Using again the asymptotic expansions of the Bessel functions one
gets $ {\rm det}[Q^{\rm TM}(R_1,R_2,\epsilon)] \propto a I_{n-m}
(\beta \epsilon / a) \frac{e^{\beta(R_2-R_1)/a}}{2\sqrt{R_1R_2}\beta}$. The
equations above can be combined to obtain
\begin{eqnarray}
M^{\rm TM}(\beta) &=&\, {\rm det}\, \left[I_{n-m}(\beta \epsilon/a)
\frac{I_m(\alpha\beta)}{I_n(\alpha\beta)} \left[1 -
(-1)^{m+n}\frac{ I_n(\beta)K_m(\alpha\beta)}{K_n(\beta)
I_m(\alpha\beta)}\right]\right ]\nonumber \\
&\times &  {\rm det}\,
I_{nm}^{-1} \left( \beta \frac{\epsilon}{a} \right),
\label{Mdfin}
\end{eqnarray}
where $ I_{nm}^{-1}(\beta \epsilon / a)$ denotes the inverse
matrix of $ I_{n-m}(\beta \epsilon /a)$ and $\alpha \equiv b/a$.
Computing explicitly the determinant one can show that the factor
$I_m(\alpha\beta) / I_n(\alpha\beta)$ cancels out. Moreover,
writing $M^{\rm TM}$ as a single determinant we get
\begin{equation}
M^{\rm TM}(\beta)={\rm det} [\delta_{np}-A_{np}^{\rm TM}],
\end{equation}
with
\begin{equation}
A_{np}^{\rm TM}=(-1)^{n}\frac{I_n(\beta)}{K_n(\beta)} \sum_m
(-1)^m \frac{K_m(\alpha\beta)}{I_m(\alpha\beta)} I_{n-m} \left(
\beta\frac{\epsilon}{a} \right) I_{mp}^{-1} \left(
\beta\frac{\epsilon}{a} \right). \label{atmfin}
\end{equation}
The addition theorem for the modified Bessel functions, ${\cal
C}_m(u\pm v)=\sum_p {\cal C}_{m\mp p}(u)J_{p}(v)$ \cite{foot},
implies that $I_{mp}^{-1}(\beta \epsilon/a)=(-1)^{m-p}
I_{m-p}(\beta \epsilon / a)$. Finally, the elements of the matrix
$A^{\rm TM}$ read
\begin{equation}
A_{np}^{\rm TM}= \frac{I_n(\beta)}{K_n(\beta)}
\sum_m \frac{K_m(\alpha\beta)}{I_m(\alpha\beta)}
I_{n-m} \left( \beta\frac{\epsilon}{a} \right) I_{p-m}\left( \beta\frac{\epsilon}{a} \right),
\label{Atm}
\end{equation}
where we omitted a global factor $(-1)^{n+p}$ because it does not contribute to the determinant.

The analysis for the TE modes is straightforward,  the main
difference being that $Q^{\rm TE}(a,b,\epsilon)$ contains
derivatives of those Bessel functions that do not depend on the
eccentricity. Therefore, following similar steps it is possible to
show that
\begin{equation}
M^{\rm TE}(\beta)={\rm det}[\delta_{np}-A_{np}^{\rm TE}],
\end{equation}
where
\begin{equation}
A_{np}^{\rm TE}= \frac{I'_n(\beta)}{K'_n(\beta)}
\sum_m \frac{K'_m(\alpha\beta)}{I'_m(\alpha\beta)}
I_{n-m}\left( \beta\frac{\epsilon}{a} \right) I_{p-m}\left( \beta\frac{\epsilon}{a} \right).
\label{Ate}
\end{equation}
The function $M$ for the electromagnetic field is the product $M=M^{\rm TE}M^{\rm TM}$,
and therefore the interaction energy is the sum of the TE and TM contributions
\begin{eqnarray}
E_{12}&=&\frac{L}{4\pi a^2}\int_0^{\infty}d\beta\, \beta\, \ln
M(\beta)=\frac{L}{4\pi a^2}\int_0^{\infty}d\beta\, \beta\, \ln
M^{\rm TE}(\beta) \nonumber \\
&+&\frac{L}{4\pi a^2}\int_0^{\infty}d\beta\, \beta\, \ln M^{\rm
TM}(\beta)=E^{{\rm TE}} + E^{{\rm TM}} . \label{e12tmte}
\end{eqnarray}

In order to calculate the exact Casimir interaction energy one
needs to perform a numerical evaluation of the determinants. We
find that as $\alpha$ approaches smaller values, larger matrices
are needed for ensuring convergence. Moreover, for increasing
values of the eccentricity $\epsilon$ it is necessary to include
more terms in the series defining the coefficients $A_{np}^{\rm
TE, TM}$. In Fig. 3 we plot the interaction energy difference
$\Delta E= E_{12} -E_{12}^{\rm cc}$ between the eccentric
($E_{12}$) and the concentric ($E_{12}^{\rm cc}$) configurations
as a function of $\alpha$ for different values of
$\delta$. As we will show below, these numerical results interpolate between the
PFA and the asymptotic behavior for large $\alpha$. Fig. 4 shows
the complementary information, with the Casimir energy as a
function of $\delta$ for various values of $\alpha$, showing
explicitly that the concentric equilibrium position is unstable.


\begin{figure}
\centerline{\psfig{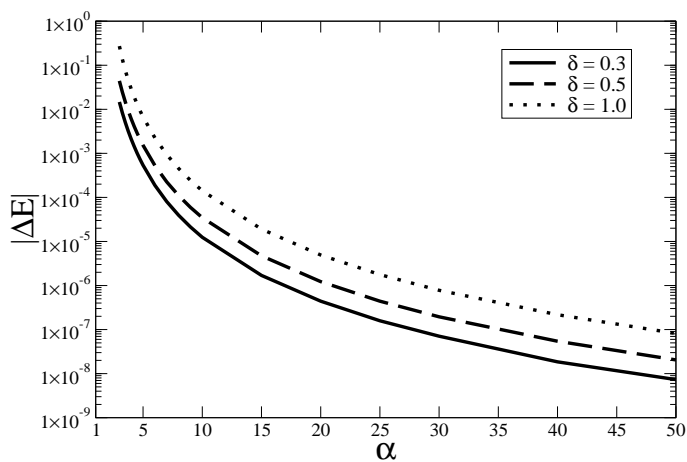}}
\caption{Exact Casimir interaction energy difference $|\Delta E|$ between the eccentric and concentric configurations as a function of $\alpha=b/a$ for different values of $\delta=\epsilon/a$.
Here $\Delta E = E_{12} - E_{12}^{\rm cc}$.
Energies are measured in units of $L/4 \pi a^2$. These results interpolate between the $(\alpha-1)^{-5}$ behavior
for $\alpha \rightarrow 1$, and the $(\alpha^4 \log \alpha)^{-1}$ behavior for $\alpha \gg 1$.}
\label{fig3}
\end{figure}


\begin{figure}
\centerline{\psfig{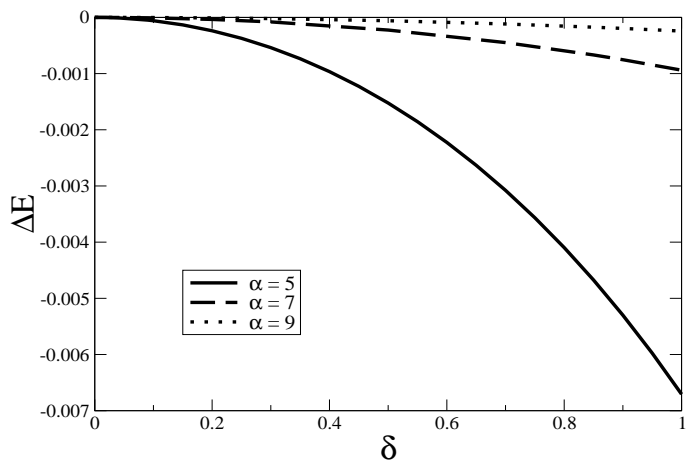}}
\caption{Exact Casimir interaction energy difference $\Delta E$ between the eccentric and concentric configurations as a function of $\delta=\epsilon/a$ for different values of $\alpha=b/a$. Energies are measured
in units of $L/4 \pi a^2$. The maximum at $\delta=0$ shows the instability of the concentric equilibrium position.}
\label{fig4}
\end{figure}


\subsection{Concentric Cylinders}

The exact Casimir interaction between concentric cylinders \cite{Mazzitelli2003} can be
obtained as a particular case of the exact formulas (\ref{Atm}), (\ref{Ate}).
In the concentric limit $\epsilon =0$, the matrices that appear in
the definition of $M^{\rm TE}$ and $M^{\rm TM}$ become diagonal and the Casimir energy reads
\cite{Mazzitelli2003}
\begin{equation}
E_{12}^{\rm cc} = {L \over 4\pi a^2} \int_{0}^{\infty} d\beta \
\beta\ln M^{\rm cc}(\beta), \label{cc}
\end{equation}
where
\begin{equation}
M^{\rm cc}(\beta)=\prod_n \left[1-{I_n(\beta)K_n(\alpha
\beta)\over I_n(\alpha \beta)K_n(\beta)}\right]
\left[1-{I'_n(\beta)K'_n(\alpha \beta) \over I'_n(\alpha
\beta)K'_n(\beta)}\right] . \label{Mcc}
\end{equation}
The first factor corresponds to Dirichlet (TM) modes and the
second one to Neumann (TE) modes.

The proximity limit $\alpha - 1\ll 1$ has already been analyzed
for the concentric case \cite{Mazzitelli2003}. In order to
compute the concentric Casimir interaction energy in this limit
it was necessary to perform the summation over all values of $n$.
As expected, the resulting value is equal to the one obtained via
the proximity approximation, namely
\begin{equation}
E_{12, {\rm PFA}}^{\rm TE, cc} = E_{12, {\rm PFA}}^{\rm TM, cc} = \frac{1}{2} E_{12, {\rm PFA}}^{\rm EM, cc} =
- \frac{\pi^3 L}{720 a^2} \; \frac{1}{(\alpha-1)^3}.
\end{equation}
Here $E^{\rm EM, cc}$ denotes the full electromagnetic Casimir interaction energy in the concentric configuration.

In the large $\alpha$ limit one can show that only the $n=0$ term contributes to the
interaction energy
\begin{equation}
E_{12}^{\rm cc} \approx  {L \over 4\pi b^2} \int_{0}^{\infty} dx \  x \left[\ln
\left(1 - \frac{I_0(\frac{x}{\alpha}) K_0(x)}{I_0(x) K_0(\frac{x}{\alpha})}\right) + \ln
\left(1 - \frac{I'_0(\frac{x}{\alpha}) K'_0(x)}{I'_0(x) K'_0(\frac{x}{\alpha})}\right)\right]. \end{equation}
Using the small argument behavior of the Bessel functions it is easy to prove
that, in the limit $\alpha \gg 1$, the TM mode contribution dominates, giving
\begin{equation} E_{12}^{\rm cc}
\approx  - {L \over 4\pi b^2\ln\alpha} \int_{0}^{\infty} dx \
x \frac{K_0(x)}{I_0(x)} \approx - {1.26 L \over 8\pi
b^2\ln\alpha}. \label{cclargea}
\end{equation}
Note that the modulus of the energy decreases logarithmically with
$\alpha$. Fig. 5 depicts the exact Casimir interaction energy
between concentric cylinders as a function of $\alpha$, for values
that interpolate between the above mentioned limiting cases.

It is worth noticing that, while for small values of $\alpha$ both
TM and TE modes contribute with the same weight to the interaction
energy, the TM modes dominate in the large $\alpha$ limit.


\begin{figure}
\centerline{\psfig{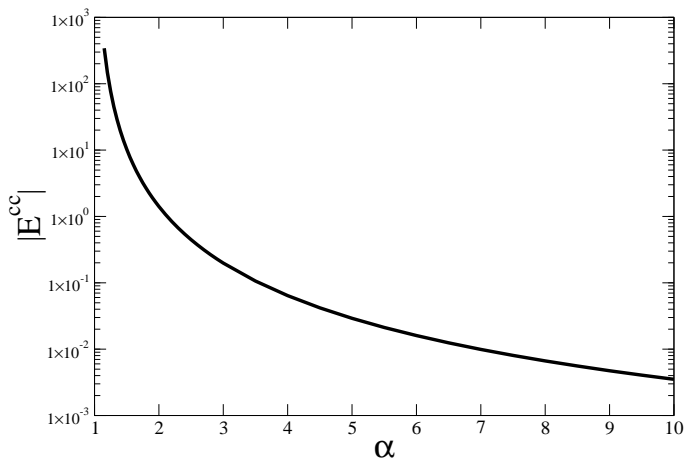}}
\caption{Modulus of the Casimir interaction energy in the
concentric case as a function of $\alpha=b/a$. Energies are
measured in units of $L/4 \pi a^2$. These results interpolate
between the $(\alpha-1)^{-3}$ behavior for $\alpha \rightarrow 1$,
and the $(\alpha^2 \log \alpha)^{-1}$ behavior for $\alpha \gg
1$.} \label{fig5}
\end{figure}

\subsection{A cylinder in front of a plane}

It is interesting to see how the Casimir energy for the
cylinder-plane configuration is contained as a particular case of
the exact formula derived in Section {\it 3.2}. To do that let us
consider a cylinder of radius $a$ in front of an infinite plane,
and let us denote by $H$ the distance between the center of the
cylinder and the plane. The eccentric cylinders formula should
reproduce the cylinder-plane Casimir energy in the limit
$b,\epsilon \rightarrow\infty$ keeping $H=b-\epsilon$ fixed (see
Fig. 1).

We note that for $x\gg h >1$ the ratio of Bessel functions appearing in Eq.(\ref{Atm}) can be approximated by
\begin{equation}
\frac{I_{m-n}(x)I_{m-p}(x)}{I_m(x+h)}\simeq I_{m-n-p}(x-h) .
\label{prodI}
\end{equation}
This is trivially true for fixed $m$ and large values of $x$, as can be seen from the
large argument expansion of the Bessel functions. Moreover, using the uniform expansion of
the Bessel functions, it can be shown that Eq.(\ref{prodI}) is also valid in the large $m$ limit,
for all values of $x$. Therefore we approximate
\begin{eqnarray}
\sum_m\frac{K_m(x+h)}{I_m(x+h)}I_{m-n}(x)I_{m-p}(x)& \simeq &  \sum_m
K_m(x+h)I_{m-p-n}(x-h) \nonumber \\ &=&K_{n+p}(2h),
\label{addition}
\end{eqnarray}
where in the last equality we used the addition theorem of Bessel
functions. Inserting this result (with $x\equiv \beta\epsilon/a$
and $h\equiv \beta H/a$) in Eq.(\ref{Atm}) we get
\begin{equation}
A_{np}^{\rm TM}\simeq \frac{I_n(\beta)}{K_n(\beta)}K_{n+p}(2\beta H/a)
\equiv A_{np}^{\rm TM, c-p},
\end{equation}
which coincides with the known result for the Dirichlet matrix elements for the cylinder-plane geometry
\cite{Emig2006,Bordag2006}. The TE modes can be analyzed in the same fashion. Using that for large $x$
\begin{equation}
\frac{K'_m(x)I_m(x)}{K_m(x)I'_m(x)}\simeq -1\,\, ,
\end{equation}
one can prove that
\begin{eqnarray}
\sum_m\frac{K'_m(x+h)}{I'_m(x+h)}I_{m-n}(x)I_{m-p}(x)& \simeq &
-\sum_m K_m(x+h)I_{m-p-n}(x-h)\nonumber \\
&=& -K_{n+p}(2h).
\label{addition2}
\end{eqnarray}
Therefore, from Eq.(\ref{Ate}) we obtain
\begin{equation}
A_{np}^{\rm TE}\simeq -\frac{I'_n(\beta)}{K'_n(\beta)}K_{n+p}(2\beta
H/a) \equiv A_{np}^{\rm TE, c-p},
\end{equation}
which is the result for the TE modes in the cylinder-plane geometry \cite{Emig2006,Bordag2006}.

\section{Quasi-concentric cylinders}

We will now consider a situation in which the eccentricity of the
configuration is much smaller than the radius of the inner
cylinder, i.e., $\delta = \epsilon /a \ll 1$. As discussed in
\cite{Dalvit2004} this configuration may be relevant for
performing null experiments to look for extra gravitational
forces. Note that we do not assume that the radius of the inner
and outer cylinder are similar, so the proximity approximation is
in general not valid for this configuration.

As described in the previous section, when $\epsilon =0$ the matrix
defining the eigenfrequencies is diagonal. When considering a
small non-vanishing eccentricity, the behavior of the Bessel
functions for small arguments $I_{m-n}(\beta \delta)\sim (\beta
\delta)^{n-m}$ suggests that one only needs to use matrix elements
near the diagonal. Using this idea, we will approximate the
Casimir interaction energy by keeping only terms proportional to $I_0,
I_1$ and $I_1^2$. In this approximation, the matrices
$\delta_{np}-A_{np}^{\rm TM}$ and $\delta_{np}-A_{np}^{\rm TE}$  become
tridiagonal matrices,  and the $\epsilon-$dependent part of the Casimir energy
will be quadratic in the eccentricity.

We will describe in detail the case of the Dirichlet modes; the treatment of Neumann modes is
similar. To order $O(\delta^2)$ the only non-vanishing elements of the matrix $A_{np}^{\rm TM}$ are
\begin{eqnarray}
A_{n,n}^{\rm TM}&\simeq &\frac{I_n(\beta)}{K_n(\beta)}\left[
\frac{K_n(\alpha\beta)}{I_n(\alpha\beta)}I_0^2(\delta\beta) +
\frac{K_{n-1}(\alpha\beta)}{I_{n-1}(\alpha\beta)}I_1^2(\delta\beta)
+\frac{K_{n+1}(\alpha\beta)}{I_{n+1}(\alpha\beta)}I_1^2(\delta\beta)\right],\nonumber\\
A_{n,n+1}^{\rm TM}&\simeq &\frac{I_n(\beta)}{K_n(\beta)}\left[
\frac{K_n(\alpha\beta)}{I_n(\alpha\beta)}
+\frac{K_{n+1}(\alpha\beta)}{I_{n+1}(\alpha\beta)}\right]I_0(\delta\beta)I_1(\delta\beta),\nonumber\\
A_{n+1,n}^{\rm TM}&\simeq
&\frac{I_{n+1}(\beta)}{K_{n+1}(\beta)}\left[
\frac{K_n(\alpha\beta)}{I_n(\alpha\beta)}
+\frac{K_{n+1}(\alpha\beta)}{I_{n+1}(\alpha\beta)}\right]I_0(\delta\beta)I_1(\delta\beta).
\label{tri}
\end{eqnarray}
We split the matrix $A^{\rm TM}$ into three terms, $A^{\rm
TM}={\cal D}^{\rm TM, cc}+{\cal D}^{\rm TM} + {\cal N}^{\rm TM}$,
where ${\cal D}^{\rm TM, cc}$ is the diagonal matrix corresponding
to the concentric case, ${\cal D}^{\rm TM}$ the diagonal part  of
the matrix that depends on $\delta$, and ${\cal N}^{\rm TM}$ is
the non-diagonal part of the matrix. The non-vanishing matrix
elements are
\begin{eqnarray}
{\cal D}_{n,n}^{\rm TM, cc} &=& \frac{I_n(\beta)}{K_n(\beta)} \frac{K_n(\alpha\beta)}{I_n(\alpha\beta)},\nonumber\\
{\cal D}_{n,n}^{\rm TM} &=&A_{n,n}^{\rm TM}-{\cal D}_{n,n}^{\rm TM, cc},\nonumber\\
{\cal N}_{n,n+1}^{\rm TM} &=&A_{n,n+1}^{\rm TM} , \quad {\cal N}_{n+1,n}^{\rm TM} =
A_{n+1,n}^{\rm TM} .
\label{tri2}
\end{eqnarray}
Note that although ${\cal D}^{\rm TM}=O(\delta^2)$ and ${\cal N}^{\rm TM}=O(\delta)$, both give quadratic
contributions
to the determinant. Up to this order we have
\begin{eqnarray}
\ln {\rm det}\,[1-A^{\rm TM}]&\simeq& \ln {\rm det}\,[1-{\cal D}^{\rm TM, cc}] +
\ln {\rm det} \left[ 1-\frac{{\cal D}^{\rm TM}}{1-{\cal D}^{\rm TM, cc}} \right]\nonumber \\ &+&
\ln {\rm det} \left[ 1-\frac{{\cal N}^{\rm TM}}{1-{\cal D}^{\rm TM, cc}} \right].
\label{dettri}
\end{eqnarray}
The first term is associated to the interaction energy between
concentric cylinders $E_{12}^{\rm cc}$, studied in Section {\it 3.3}, and
being $\delta$-independent does not
contribute to the force between eccentric cylinders.
The second term can be easily evaluated
\begin{equation}
\ln {\rm det} \left[ 1-\frac{{\cal D}^{\rm TM}}{1-{\cal D}^{\rm
TM, cc}} \right] \simeq\ln\left( 1-{\rm tr}\, \frac{{\cal D}^{\rm
TM}}{1-{\cal D}^{\rm TM, cc}}\right)\simeq -\sum_n\frac{{\cal
D}^{\rm TM}_{n,n}}{1-{\cal D}_{n,n}^{\rm TM, cc}} . \label{diagpart}
\end{equation}
To compute the last term in  Eq.(\ref{dettri}) we use that the
determinant of an arbitrary tridiagonal matrix $T$ of dimension
$p$ can be calculated using the recursive relation
${\rm det} [ T_{ \{ p\} } ] = T_{p,p}  {\rm det} [T_{ \{ p-1\} } ] -
T_{p,p-1} T_{p-1, p} \, {\rm det}[ T_{ \{ p-2\} } ]$,
where $T_{\{ k\} }$ denotes the submatrix formed by the first $k$
rows and columns of $T$. Up to quadratic order in $\delta$ we
obtain
\begin{eqnarray}
\ln {\rm det} \left[  1-\frac{{\cal N}^{\rm TM}}{1-{\cal D}^{\rm
TM, cc}} \right] &\simeq &\ln\left( 1- \sum_n \frac{A_{n, n+1}^{\rm
TM} \; A_{n+1, n}^{\rm TM}}{(1-{\cal D}_{n,n}^{\rm TM, cc})(1-{\cal
D}_{n+1,n+1}^{\rm TM, cc})} \right) \nonumber \\ &\simeq & -\sum_n \frac{A_{n,
n+1}^{\rm TM} \; A_{n+1, n}^{\rm TM}}{(1-{\cal D}_{n,n}^{\rm TM,
cc})(1-{\cal D}_{n+1,n+1}^{\rm TM, cc})}. \label{nondiagpart}
\end{eqnarray}
Putting all together, the TM part of the Casimir interaction energy between quasi-concentric
cylinders can be written
as
\begin{equation}
E_{12}^{\rm TM} = E_{12}^{{\rm TM, cc}} - \frac{L \epsilon^2}{4
\pi a^4} \sum_n  \int_0^{\infty} d\beta \; \beta^3
\frac{1}{1-{\cal D}^{\rm TM, cc}_{n,n}} \left[ {\cal D}^{\rm TM}_n +
\frac{{\cal N}^{\rm TM}_n}{1-{\cal D}^{\rm TM, cc}_{n+1,n+1}} \right].
\label{finalqcc}
\end{equation}
Here
\begin{eqnarray}
{\cal D}^{\rm TM}_n & \equiv & \frac{ {\cal D}^{\rm TM, cc}_{n,n}}{2} + \frac{I_n(\beta)}{4 K_n(\beta)} \left[
\frac{K_{n-1}(\alpha\beta)}{I_{n-1}(\alpha \beta)} +
\frac{K_{n+1}(\alpha \beta)}{I_{n+1}(\alpha \beta)} \right] , \nonumber \\
{\cal N}^{\rm TM}_n & \equiv & \frac{I_n(\beta) I_{n+1}(\beta)}{4
K_n(\beta) K_{n+1}(\beta)} \left[ \frac{K_{n}(\alpha
\beta)}{I_{n}(\alpha \beta)} + \frac{K_{n+1}(\alpha
\beta)}{I_{n+1}(\alpha \beta)} \right]^2 . \label{newd}
\end{eqnarray}
The corresponding formulas for the TE modes can be obtained from
these ones by replacing the Bessel functions by their derivatives
with respect to the argument.

The expression for the Casimir energy for quasi-concentric
cylinders derived in this section is far simpler than the exact
formulas Eqs.(\ref{Atm}), (\ref{Ate}). It is very useful for the
analytical and numerical evaluation of the Casimir energy in the
different limiting cases we will study below: the large distance
limit ($a \ll b$), for which one obtains a logarithmic decay of
the energy, and the small distance limit ($a \simeq b$), where the
proximity approximation holds. The first case is very simple to
handle because the energy is dominated by the lowest modes, while
the second case is much more involved.

\subsection{Large distances: logarithmic decay}

When the ratio of the outer and the inner  radii $\alpha = b/a$ is
much larger than one, the exact Casimir energy is dominated by the
lowest term $n=0$ in the summation. Moreover, it can be shown that
the contribution of the Dirichlet modes is much larger than the
contribution of the Neumann modes. Therefore, from
Eq.(\ref{e12tmte}) we get, in the limit $\alpha \rightarrow
\infty$,
\begin{equation}
E_{12}^{\infty} \simeq {L \over 4\pi a^2} \int_{0}^{\infty} d\beta
\; \beta \ln(1-A_{00}^{\rm TM}(\beta)) \simeq -{L \over 4\pi
a^2\alpha^2}\int_{0}^{\infty} dx \; x \; A_{00}^{\rm TM} \left(
\frac{x}{\alpha} \right) , \label{large}
\end{equation}
where
\begin{equation} A_{00}^{\rm TM}\left( \frac{x}{\alpha}\right) \simeq
\frac{I_0(\frac{x}{\alpha})}{K_0(\frac{x}{\alpha})}\left[
\frac{K_0(x)}{I_0(x)}I_0^2\left(\frac{\delta x}{\alpha}\right) +
2\frac{K_1(x)}{I_1(x)}I_1^2\left(\frac{\delta
x}{\alpha}\right)\right]\, .
\end{equation}
Using the small argument expansion of the Bessel functions it is
easy to see that
\begin{equation}
A_{00}^{\rm TM}\left(\frac{x}{\alpha} \right)\simeq
\frac{1}{\ln\alpha}\left[\frac{K_0(x)}{I_0(x)}+
\frac{\delta^2x^2}{2\alpha^2}\left(\frac{K_0(x)}{I_0(x)}+\frac{K_1(x)}{I_1(x)}\right)\right].
\label{a00}
\end{equation}
In this expression, valid when $a,\epsilon \ll b$, we
kept the leading terms proportional to $(\alpha^2\ln\alpha )^{-1}$
and only the subleading terms that depend on the eccentricity.
Inserting Eq.(\ref{a00}) into Eq.(\ref{large}) and computing
numerically the integrals we find
\begin{equation}
E_{12}^{\infty} \simeq  -{L \over 8\pi b^2\ln\alpha}
\left(1.26+3.33 \frac{\epsilon^2}{b^2} \right),
\label{resultlarge}
\end{equation}
where the first term is the concentric contribution
$E_{12}^{\infty, {\rm cc}}$ derived before (see
Eq.(\ref{cclargea})). It is worth to note that Eqs. (\ref{a00})
and (\ref{resultlarge}) have been derived under the assumption
$\ln\alpha\gg 1$, and therefore are valid for extremely large
values of $\alpha$. For intermediate values $\alpha\gg 1,
\ln\alpha =O(1)$, the interaction energy is also dominated by the
Dirichlet $n=0$ term. The final result is still of the form given
in Eq.(\ref{resultlarge}), with numerical coefficients that depend
logarithmically on $\alpha$.  In Fig. 6 we plot the ratio between
the exact Casimir interaction energy difference $\Delta E \equiv
E_{12} - E_{12}^{\rm cc}$ and its asymptotic expression $\Delta
E_{\infty} \equiv  E_{12}^{\infty} - E_{12}^{\infty,{\rm cc}}$ as
a function of $\alpha$. As mentioned, extremely large values of
$\alpha$ are needed in order for the ratio of energies to
asymptotically approach 1. From Eq.(\ref{resultlarge}) we see that
the force between cylinders in the limit $a,\epsilon\ll b$ is
proportional to $L \epsilon /b^4\ln(b/a)$. The weak logarithmic
dependence on the ratio $b/a$ is characteristic of the cylindrical
geometry (see also \cite{Emig2006,Bordag2006}), and it is also
found in the electrostatic counterpart of the Casimir energy, that
we briefly analyze next.


\begin{figure}
\centerline{\psfig{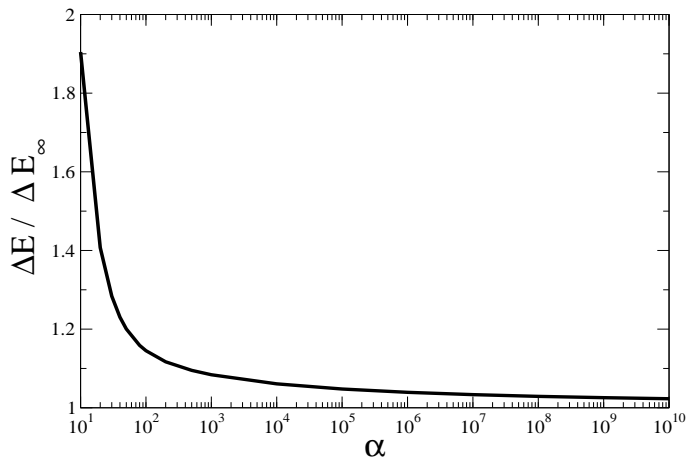}}
\caption{Ratio of the exact $\Delta E$ and asymptotic $\Delta E_{\infty}$ Casimir energy differences in the limit of small eccentricity $\epsilon \ll a$.
In the $\alpha \rightarrow \infty$ limit, the Casimir energy difference between eccentric
and concentric configurations decays logarithmically as $(\alpha^4 \log \alpha)^{-1}$.}
\label{fig6}
\end{figure}

The electrostatic capacity for the system of two eccentric cylinders is given by
\begin{equation}
C=\frac{2\pi\epsilon_0L}{\ln [Y+\sqrt{Y^2-1}]}, \label{capacity}
\end{equation}
where $Y=(a^2+b^2-\epsilon^2)/2ab$, and $\epsilon_0$ is the permittivity of vacuum.
Therefore, the electrostatic force between cylinders kept at a fixed potential difference $V$ is
\begin{equation}
F_{\rm elec}=\frac{\epsilon}{ab}\frac{\pi\epsilon_0 V^2L}{\sqrt{Y^2-1}\ln^2[Y+\sqrt{Y^2-1}]} .
\label{electroforce}
\end{equation}
In the quasi-concentric case we can set $\epsilon =0$ in the
definition of $Y$. In the large $\alpha$ limit  we get
\begin{equation}
F_{\rm elec} \simeq \frac{L \epsilon}{b^2\log ^2 \left( \frac{b}{a} \right) } .
\end{equation}
Just as in the Casimir case, in the limit $a\ll b$ the Coulomb force decays logarithmically with
the ratio $a/b$.

\subsection{Small distances: the proximity approximation}

The proximity limit for concentric cylinders has been reviewed in
Sec. {\it 3.3}; the case of a cylinder in front of a plane has
been considered in detail in \cite{Bordag2006}. In this section we
extend these results to the case of quasi-concentric cylinders.

We will concentrate on calculating the Casimir interaction energy
difference $\Delta E \equiv E_{12} - E_{12}^{\rm cc}$ between the
eccentric ($E_{12}$) and the concentric ($E_{12}^{\rm cc}$)
configurations. As the small distance limit is dominated by the
large-$n$ modes, the key point in the derivation of the PFA from
the exact expression of the Casimir energy is the use of the
uniform approximation for the Bessel functions. In the large $n$
limit, and to leading order in $\alpha -1$ one has
\begin{eqnarray}
\frac{I_n(\beta)}{K_n(\beta)}\frac{K_n(\alpha\beta)}{I_n(\alpha\beta)}&
\simeq & e^{-2n(\alpha-1)h(x)},\nonumber \\
\frac{I_n(\beta)}{K_{n}(\beta)}\frac{K_{n\pm
1}(\alpha\beta)}{I_{n\pm 1}(\alpha\beta)}& \simeq &
e^{-2n(\alpha-1)h(x)}\left[\frac{1+h(x)}{x}\right]^{\pm 2},
\label{unifapprox}
\end{eqnarray}
where $\beta = n x$ and $h(x)=\sqrt{1+x^2}$. Inserting  these
approximations in Eq.(\ref{newd}) we get
\begin{equation}
{\cal D}^{\rm TM}_n(n x)=\frac {e^{-2 n (\alpha-1)h(x)}}{2} \left[
1+ \frac{1}{2} \left( \frac{1+h(x)}{x} \right)^{2} +
\frac{1}{2}\left( \frac{1+h(x)}{x} \right)^{-2} \right].
\end{equation}

\begin{figure}
\centerline{\psfig{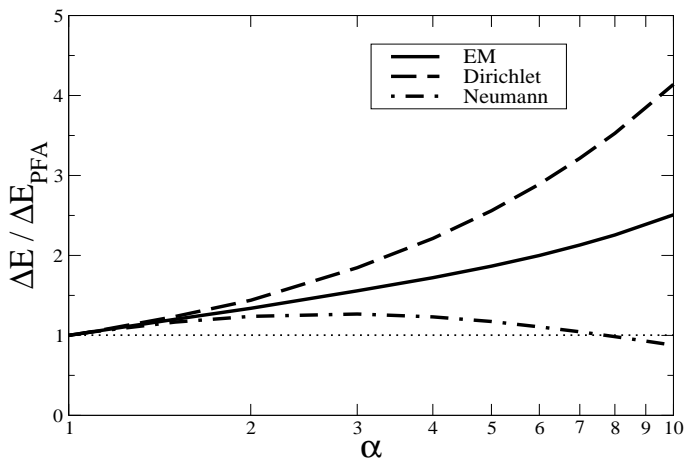}}
\caption{Ratio of the exact and PFA Casimir interaction energy differences $\Delta E = E_{12} -E_{12}^{\rm cc}$
between eccentric ($E_{12}$) and concentric ($E_{12}^{\rm cc}$) cylinders in the limit of small eccentricity
$\epsilon \ll a$. The curve EM denotes the full electromagnetic Casimir energy.}
\label{fig7}
\end{figure}

The contribution to the interaction energy coming from the
diagonal part of the matrix can be written as (see
Eq.(\ref{finalqcc}))
\begin{equation}
\Delta E_{\cal D}^{\rm TM}=-\frac{L\delta^2}{2\pi a^2}
\sum_{n,k\geq 1} \int_0^\infty d\beta \; \beta^3 \; {\cal D}^{\rm
TM}_n ({\cal D}_{n,n}^{\rm TM, cc})^{k-1} , \label{ED}
\end{equation}
where we replaced the sum over all integers $n$ by twice the sum
over the positive integers (the term $n=0$ gives a subleading
contribution for small $\alpha -1$). Inserting the uniform expansions
into Eq.(\ref{ED}) and changing variables in the integral we obtain
\begin{eqnarray}
\Delta E^{\rm TM}_{\cal D}&=& -\frac{L\delta^2}{4\pi a^2}\sum_{n,k\geq 1}n^4\int_0^\infty
dx \;  x^3 e^{-2n(\alpha-1)h(x)k} \nonumber \\
&& \times \left[ 1+\frac{1}{2}\left( \frac{1+h(x)}{x} \right)^{2}+
\frac{1}{2}\left( \frac{1+h(x)}{x} \right)^{-2} \right] .
\label{ED2}
\end{eqnarray}
To leading order in $\alpha -1$ the sum over $n$ gives
$\sum_{n} n^4 e^{-2n(\alpha-1)h(x)}\simeq 24/ [2 h(x)(\alpha-1)k]^5$.
Next we perform first the sum over $k$ and then the integral over
$x$. We get
\begin{equation}
\Delta E^{\rm TM}_{\cal D} = -\frac{3}{8}\frac{L\delta^2\zeta(5)}{\pi a^2 (\alpha - 1)^5} ,
\label{ED3}
\end{equation}
where $\zeta(x)$ is the Riemann zeta function.

The evaluation of the non diagonal contribution to the Casimir
energy can be done using  similar steps, starting from
Eq.(\ref{nondiagpart}). In the proximity limit, we can approximate
${\cal D}_{n+1,n+1}^{\rm TM, cc}$ by ${\cal D}_{n,n}^{\rm TM, cc}$ in
the denominator of that equation. Therefore, using
Eq.(\ref{finalqcc})we write the non diagonal contribution to the
energy as
\begin{equation}
\Delta E^{\rm TM}_{\cal ND} \simeq -\frac{L\delta^2}{2\pi
a^2}\int_0^\infty d\beta \; \beta^3 \sum_{n,k\geq 1}  {\cal N}^{\rm TM}_n
 \;k \; ( {\cal D}_{n,n}^{\rm
TM, cc})^{k-1}. \label{END2}
\end{equation}
Now we use the uniform expansion for the Bessel functions in
Eq.(\ref{newd}) to obtain
\begin{eqnarray}
{\cal N}^{\rm TM}_n &\simeq & \frac{e^{-4n(\alpha -1)h(x)}}{2} \nonumber \\
&& \times \left[ 1+\frac{1}{2}\left( \frac{1+h(x)}{x} \right)^{2}
+ \frac{1}{2}\left( \frac{1+h(x)}{x} \right)^{-2} \right]
\label{an+1} .
\end{eqnarray}
Replacing Eq.(\ref{an+1}) into Eq.(\ref{END2}) we get
\begin{eqnarray}
\Delta E^{\rm TM}_{\cal ND} &=& -\frac{L\delta^2}{8\pi a^2}\sum_{n,k\geq1} n^4 \int_0^{\infty} dx \;  x^3 \;
e^{-2n(k+1)(\alpha-1)h(x)}\; k \nonumber \\
&&
\times \left[
2 + \left( \frac{1+h(x)}{x} \right)^{2} + \left( \frac{1+h(x)}{x} \right)^{-2}
\right].
\end{eqnarray}
As before, we first compute the sum over $n$, and expand the
result to leading order in $\alpha-1$. The sum over $k$ can be
calculated using that $\sum_{k\geq 1}\frac{k}{(k+1)^5}=\zeta (4)-\zeta(5)$.
Finally, we compute analytically the remaining integrals to get
\begin{equation}
\Delta E^{\rm TM}_{\cal ND}=-\frac{3}{8}\frac{L\delta^2}{\pi a^2(\alpha-1)^5}(\zeta(4)-\zeta(5)).
\end{equation}
The contribution of the Dirichlet modes to the Casimir interaction
energy in the limit $\alpha \rightarrow 1$ is therefore
\begin{equation}
\Delta E^{\rm TM}=\Delta E^{\rm TM}_{\cal D} + \Delta E^{\rm TM}_{\cal ND}=
-\frac{L\delta^2}{a^2(\alpha-1)^5}\frac{\pi^3}{240}.
\end{equation}
It can be shown that the contribution of the TE modes to the
interaction energy in the short distance limit is equal to that of
the TM modes, as expected from the parallel plate configuration.
Indeed, the uniform expansion for the ratio of Bessel functions is
equal to the expansion for the derivatives, i.e.,
\begin{eqnarray}
\frac{I_n'(\beta)}{K_n'(\beta)}\frac{K_n'(\alpha\beta)}{I_n'(\alpha\beta)}&
\simeq & e^{-2n(\alpha-1)h(x)},\nonumber\\
\frac{I_n'(\beta)}{K_{n}'(\beta)}\frac{K_{n\pm
1}'(\alpha\beta)}{I_{n\pm 1}'(\alpha\beta)}& \simeq &
e^{-2n(\alpha-1)h(x)}\left[\frac{1+h(x)}{x}\right]^{\pm 2},
\label{unifapprox2}
\end{eqnarray}
and therefore all calculations can be repeated without changes.

The final result for the Casimir interaction energy difference in the small distance approximation is
\begin{equation}
\Delta E_{\rm PFA}^{\rm TE} = \Delta E_{\rm PFA}^{\rm TM} = \frac{1}{2} \Delta E_{\rm PFA}^{\rm EM} =
- \frac{\pi^3 L \epsilon^2}{240 a^4 (\alpha-1)^5} ,
\end{equation}
where $\Delta E^{\rm EM}$ denotes the full electromagnetic Casimir
energy difference between eccentric and concentric configurations.
Fig. 7 depicts the ratio of the exact Casimir energy difference
$\Delta E$ and the PFA limit for the almost concentric cylinders
configuration. As evident from the figure, PFA agrees with the
exact result at a few percent level only for $\alpha$ close to
unity, and then it noticeably departs from the PFA prediction. The
resulting PFA expression for the attractive Casimir force between
quasi-concentric cylinders reads
\begin{equation}
F^{\rm PFA}= \frac{\pi^3}{60}\frac{\epsilon L}{a^4(\alpha -1)^5},
\end{equation}
that reproduces the result previously obtained in \cite{Dalvit2004}.

\section{Conclusions}

We have derived an exact formula for the Casimir interaction energy
between eccentric cylinders using a mode summation technique. This formula
is written as an integral of the determinant of an infinite dimensional matrix, and it
reproduces as a particular case the interaction energy between concentric
cylinders, and as a limiting case the energy in the cylinder-plane
geometry. In the quasi-concentric case, the infinite dimensional
matrix becomes tridiagonal, and hence much easier to deal with
than the exact formula when performing analytic and numerical calculations.
We have carried out the numerical evaluation of the Casimir interaction
energy using both  the exact and tridiagonal formulas, and studied different
limiting cases of relevance for Casimir force measurements.

The large and small distance limits were analyzed. In the former case,
the Casimir energy is dominated by the lowest modes, and shows a weak
logarithmic decay, typical of cylindrical geometries. In the latter case,
the Casimir energy is dominated by the highest modes, and the exact formula
reproduces the proximity approximation. We found that the first order correction
($\alpha -1 \ll 1$) to PFA for the quasi-concentric cylinders has the form
$\Delta E / \Delta E_{\rm PFA} = 1 + s (\alpha-1) + O((\alpha-1)^2)$, where the coefficient
of the linear curvature correction is positive, $s>0$, both for TE and TM modes.
This contrasts with the first order corrections to PFA in the cylinder-plane configuration,
where the linear curvature correction to TM modes is positive, while the one for TE
modes is negative \cite{Bordag2006}.

The exact Casimir force computed in this paper, in particular for the
quasi-concentric configuration,  offers a qualitatively different
approach for implementing new experiments to measure the Casimir force and to
search for extra-gravitational forces in the micrometer and nanometer scales,
since it opens the possibility of measuring the derivative of the force using (Cavendish-like) null experiments.

\ack
We are grateful to R. Onofrio and J. Von Stecher for fruitful
discussions. We thank A. L\'opez D\'avalos for pointing
Ref.\cite{Balseiro1950} to us. The work of F.C.L. and F.D.M. is
supported by UBA, Conicet and ANPCyT (Argentina).


\begin{thebibliography}{[99]}
\bibitem{foot} This is a particular case of Eq.(\ref{addition1}), valid for $\vert v\vert < \vert u\vert$.
\end{thebibliography}
\end{document}